\documentstyle[12pt]{article}

\newcommand{\ket}{\rangle}
\newcommand{\bra}{\langle}
\begin{document}
\setlength{\baselineskip}{8mm}
\setlength{\jot}{3mm}
\begin{center}
{\Large\bf 
Linear Canonical Transformations and the Hamilton-Jacobi Theory 
in Quantum Mechanics}
\vspace{5cm}

Akihiro Ogura\footnote{ogu@mascat.nihon-u.ac.jp} 
and Motoo Sekiguchi${}^{\dagger}$\footnote{motoo@kokushikan.ac.jp}
 
\hspace{1cm}

Laboratory of Physics, Nihon University, Chiba 271-8587, Japan

and 

${}^{\dagger}$Faculty of Engineering, Kokushikan University, 
       Tokyo 154-8515, Japan

\vspace{1.5cm}

\begin{abstract}
We investigate two methods of constructing a solution of the 
Schr\"{o}dinger equation from the canonical transformation 
in classical mechanics. 
One method shows that we can formulate the solution of the 
Schr\"{o}dinger equation from 
linear canonical transformations, the other focuses on the 
generating function which satisfies the Hamilton-Jacobi equation 
in classical mechanics. 
We also show that these two methods lead to the same solution 
of the Schr\"{o}dinger equation. 
\end{abstract}
\end{center}

\newpage
\section{ Introduction }

The idea of canonical transformations provides us with a powerful 
technique for solving mechanical problems. Not only does this 
enable us to solve classical problems 
but it also gives us a clue to the quantization of classical systems. 
In paticular, the Hamilton-Jacobi theory provides the key for 
connecting classical and quantum mechanics. 

However, there are few problems in which a solution to the 
Schr\"{o}dinger equation can be found by a canonical transformation 
or the Hamilton-Jacobi theory in classical mechanics. 
One of the reasons may be the restriction of the 
non-commutativity of physical variables in quantum systems. 
The analogy of the canonical transformation in classical 
mechanics with the unitary transformation in quantum mechanics 
was pointed out by Dirac~\cite{dirac1} just after the birth 
of the quantum mechanics in 1926. 
Originally, he considered the transformation function 
$\bra q|P;t \ket$ between the old 
position $q$ and new momentum $P$ with the corresponding generating 
function $W_{2}(q, P, t)$ 
\begin{equation}
\bra q|P;t \ket = \exp\left[ \frac{i}{\hbar} W_{2}(q, P, t) \right] .
\label{eq:type2}
\end{equation}
In his next papers~\cite{dirac2, dirac3}, he changed the variable from $P$ 
to the new position $Q$ and constructed the transformation function 
\begin{equation}
\bra q|Q;t \ket = \exp\left[ \frac{i}{\hbar} W_{1}(q, Q, t) \right] .
\label{eq:type1}
\end{equation}
This form of the transformation function inspired Feynman to invent 
the celebrated path integrals~\cite{feynman}.

Recently, some authors~\cite{lee, kim, ghandour} have reconsidered the 
validity of these transformation functions (\ref{eq:type1}) for solving 
the quantum mechanical problems. 
Especially, Lee and l'Yi~\cite{lee} derived a solution of the 
time-independent Schr\"{o}dinger equation in the transformed $Q$-space, 
which was then transformed to the original $q$-space by the 
transformation function (\ref{eq:type1}). 
The relationship with the Hamilton-Jacobi theory has been investigated 
by Kim and Lee~\cite{kim}. 

In this paper, we show two methods for constructing solutions 
of the Schr\"{o}dinger equation. 
One method is that by which we construct the transformation function 
$\bra q|Q;t \ket$ 
from linear canonical transformations in classical mechanics. 
The other method is closely related to the Hamilton-Jacobi theory. 
Using the generating function $W_{1}(q,Q;t)$ which satisfies 
the Hamilton-Jacobi 
equation, we construct the solution of the Schr\"{o}dinger equation. 
It is shown that these two method leads to identical solutions. 

This paper is organized in the following way. In section 2, 
we construct the quantum transformation function from linear 
canonical transformations in classical mechanics. 
In section 3, the solution of the Schr\"{o}dinger equation is 
obtained with the aid of the Hamilton-Jacobi theory in classical 
mechanics. In section 4, we will show some applications to 
ideal systems. Section 5 is devoted to a discussion.

\section{ Linear canonical transformations 
and their transformation function }

We consider the following linear canonical transformations; 
\begin{equation}
  \left\{
    \begin{array}{@{\,}ll}
        Q(t) &=a(t) q + b(t) p  \\
        P(t) &=c(t) q + d(t) p ,
    \end{array}
  \right.
\label{eq:lct}
\end{equation}
where $(q, p)$ and $(Q, P)$ describe the old and new canonical 
variables. We assume that the coefficients $a(t)$, $b(t)$, $c(t)$ 
and $d(t)$ are all real functions of time $t$. 
In order for these linear tranformations to be canonical 
transformations, the 
Poisson bracket $[A, B]_{c}$ should be satisfied. Thus, 
the coefficients $a(t)$, $b(t)$, $c(t)$ and $d(t)$ are 
constrained; 
\begin{equation}
[Q(t), P(t)]_{c}=\frac{\partial Q(t)}{\partial q}
                 \frac{\partial P(t)}{\partial p}
                -\frac{\partial P(t)}{\partial q}
                 \frac{\partial Q(t)}{\partial p}
=a(t)d(t)-b(t)c(t)=1.
\label{eq:pb}
\end{equation}

Now, we construct the transformation function from the canonical 
transformations in classical mechanics. 
The canonical variables in eq.(\ref{eq:lct}) are raised to quantum 
numbers: we attach $\hat{~}$ to the variables to distinguish 
the q-numbers from c-numbers, 
\begin{equation}
  \left\{
    \begin{array}{@{\,}ll}
        \hat{Q}(t) &=a(t) \hat{q} + b(t) \hat{p}  \\
        \hat{P}(t) &=c(t) \hat{q} + d(t) \hat{p} . 
    \end{array}
  \right.
\label{eq:lqt}
\end{equation}
With the condition (\ref{eq:pb}) and the commutation relation 
for the old variables 
\begin{equation}
\left[\hat{q}, \hat{p} \right] 
 = \hat{q}\hat{p}-\hat{p}\hat{q} = i\hbar, 
\end{equation}
we have the commutation relation for the new variables 
\begin{equation}
  \left[ \hat{Q}(t), \hat{P}(t) \right] = i \hbar .
\label{eq:commu}
\end{equation}

We proceed by constructing a transformation 
function~\cite{moshinsky}. Firstly, 
we define the eigenstate $|Q;t \ket$ of the operator 
$\hat{Q}(t)$ with eigenvalue $Q$ to form the $Q$-representation:
\begin{equation}
  \hat{Q}(t)|Q;t \ket = Q|Q;t \ket. 
\label{eq:eigeneq}
\end{equation}

Secondly, the $q$-representation for the eigenstate $|Q;t \ket$ 
can be obtained from the differential equation 
\begin{equation}
  \bra q |\hat{Q}(t)|Q;t \ket 
  =\left\{ a(t)q -i\hbar b(t) 
          \frac{\partial}{\partial q} \right\}\bra q|Q;t \ket
  = Q \bra q|Q;t \ket. 
\label{eq:diff}
\end{equation}
Integrating (\ref{eq:diff}) with the normalization condition 
\begin{equation}
\bra Q;t|Q';t \ket = \delta(Q-Q'), 
\end{equation}
we have 
\begin{equation}
  \bra q|Q;t \ket = \Phi(Q)\times \sqrt{\frac{1}{2\pi (-i) \hbar b(t)}}
   \exp\left[ \frac{i}{\hbar}\frac{2qQ-a(t)q^{2}}{2b(t)} \right]
\label{eq:tra0}
\end{equation}
where $\Phi(Q)$ is an arbitrary phase factor which depends on $Q$. 
Here, the meaning of $(-i)$ in the square root is clarified later. 
Note that hereafter we consider the $b(t) \neq 0$ case so as 
not to lose generality in our discussion.

Thirdly, we consider the matrix element of the operator $\hat{P}(t)$ 
with an arbitrary state $|\psi \ket$, 
\begin{eqnarray}
\bra Q;t|\hat{P}(t)|\psi \ket 
&=& \int dq \bra Q;t|q\ket \bra q|\hat{P}(t)|\psi \ket \\ 
&=& \int dq \bra Q;t|q\ket \left\{ c(t)q - i\hbar d(t) 
       \frac{\partial}{\partial q} \right\} \bra q|\psi \ket .
\end{eqnarray}
Integrating by parts and using the $q$-derivative of 
eq.(\ref{eq:tra0}), we have 
\begin{equation}
\bra Q;t|\hat{P}(t)|\psi \ket 
= \left\{ \frac{d(t)}{b(t)}Q -i\hbar\frac{\partial}{\partial Q} 
   + \frac{i\hbar}{\Phi^{\ast}}
     \frac{\partial \Phi^{\ast}}{\partial Q} \right\}\bra Q;t|\psi \ket.
\end{equation}
If we set 
\begin{equation}
\Phi^{\ast}(Q) = \exp\left[ \frac{i}{\hbar} 
                            \frac{d(t)}{2b(t)} Q^{2} \right], 
\end{equation}
$\bra Q;t|\hat{P}|\psi \ket$ is given by 
\begin{equation}
  \bra Q;t|\hat{P}(t)|\psi \ket 
              = -i\hbar \frac{\partial}{\partial Q} \bra Q;t|\psi \ket, 
\label{eq:prep}
\end{equation}
so that the $q$-representation of the eigenstate $|Q;t \ket$ turns out 
to be 
\begin{equation}
\bra q|Q;t \ket = \sqrt{\frac{-1}{2\pi i \hbar b(t)}} 
     \exp\left[ \frac{i}{\hbar} \frac{2qQ-a(t)q^{2}-d(t)Q^{2}}
                                     {2b(t)} \right] .
\label{eq:trans}
\end{equation}

This is the first result in this paper. We have 
the linear canonical transformation (\ref{eq:lct}) which is 
neither a point transformation, nor a transformation from 
position to momentum space. Next, all variables are raised to 
q-number variables and we take the old and new positions 
as bases and form the transformation function 
$q \Leftrightarrow Q$. 

It is noted that 
the exponential factor, except for $(i/\hbar)$, 
\begin{equation}
W_{1}(q, Q, t) = \frac{qQ}{b(t)} - \frac{a(t)}{2b(t)}q^{2}
                                 - \frac{d(t)}{2b(t)}Q^{2}
\label{eq:gf1}
\end{equation}
is a type-1 generating function which causes a canonical 
transformation (\ref{eq:lct}) according to the ordinary 
prescription: 
\begin{equation}
p =   \frac{\partial}{\partial q} W_{1}(q,Q,t), \hspace{1cm}
P = - \frac{\partial}{\partial Q} W_{1}(q,Q,t) .
\end{equation}
The other types of generating functions are given by Legendre 
transformations, 
\begin{eqnarray}
W_{2}(q, P, t) &=& \frac{qP}{d(t)} - \frac{c(t)}{2d(t)}q^{2}
                                 + \frac{b(t)}{2d(t)}P^{2} \\
W_{3}(p, Q, t) &=& -\frac{pQ}{a(t)} + \frac{b(t)}{2a(t)}p^{2}
                                 - \frac{c(t)}{2a(t)}Q^{2} \\
W_{4}(p, P, t) &=& -\frac{pP}{c(t)} + \frac{d(t)}{2c(t)}p^{2}
                                 + \frac{a(t)}{2c(t)}P^{2} , 
\end{eqnarray}
while the transformation functions are given by Fourier 
transformations~\cite{kim}, 
\begin{eqnarray}
\bra q|P;t \ket &=& \sqrt{\frac{1}{2\pi \hbar d(t)}} 
              \exp\left[\frac{i}{\hbar}W_{2}(q, P, t)\right] \\
\bra p|Q;t \ket &=& \sqrt{\frac{1}{2\pi \hbar a(t)}} 
              \exp\left[\frac{i}{\hbar}W_{3}(p, Q, t)\right] \\
\bra p|P;t \ket &=& \sqrt{\frac{1}{2\pi i\hbar c(t)}} 
              \exp\left[\frac{i}{\hbar}W_{4}(p, P, t)\right] .
\end{eqnarray}

\section{ Schr\"{o}dinger equation and Hamilton-Jacobi theory }

In this section, we find a solution to the Schr\"{o}dinger 
equation 
\begin{equation}
  i\hbar \frac{\partial}{\partial t} \psi(q,t) = 
   \left[ -\frac{\hbar^{2}}{2m}\frac{\partial^{2}}{\partial q^{2}}
          + V(q) \right] \psi(q,t) ,
\label{eq:se}
\end{equation}
with the aid of the Hamilton-Jacobi theory in classical mechanics. 

Let us write the wave function as 
\begin{equation}
  \psi(q,t) = \exp\left[ \frac{i}{\hbar} S(q,t) \right], 
\label{eq:solu}
\end{equation}
and putting this into the above Schr\"{o}dinger equation 
(\ref{eq:se}), we obtain 
\begin{equation}
  \frac{1}{2m}\left(\frac{\partial S}{\partial q}\right)^{2} 
  + V(q) + \frac{\partial S}{\partial t} 
  -\frac{i\hbar}{2m}\frac{\partial^{2}S}{\partial q^{2}} = 0. 
\label{eq:HJ-0}
\end{equation}
This form of the equation has been well studied in the classical 
limit using the WKB formalism where the focus is on the 
stationary-state solution. 
Here, we consider this equation from a different point of view. 
We define the new function 
\begin{equation}
  F \equiv \frac{1}{2m}\frac{\partial^{2}S}{\partial q^{2}}. 
\label{eq:defF}
\end{equation}
If the function $S$ is given by a polynomial with respect to $q$ 
up to the second order, $F$ is independent of $q$ and depends 
only on $t$. 
We consider this case hereafter. Under this condition, 
the function $S$ is divided into two parts; one depending only 
on time $t$, and the other $W(q, t)$ depending on $q$ and $t$, 
\begin{equation}
  S(q,t) = W(q,t) + i\hbar \int^{t} dt' F(t'). 
\label{eq:defS}
\end{equation}
Putting this back into eq.(\ref{eq:HJ-0}), we obtain 
\begin{equation}
  \frac{1}{2m}\left(\frac{\partial W}{\partial q}\right)^{2} 
  + V(q) + \frac{\partial W}{\partial t} = 0 . 
\label{eq:HJ}
\end{equation}
This is the Hamilton-Jacobi equation that appears 
in classical mechanics. This indicates the possibility of obtaining 
a solution to the Schr\"{o}dinger equation from the 
Hamilton-Jacobi theory in classical mechanics. 
The procedure for this is as follows. 
\begin{enumerate}
\item Once we have the function $W$ which satisfies the Hamilton-Jacobi 
      equation (\ref{eq:HJ}),
\item we calculate the function $F$ from eq.(\ref{eq:defF}). 
      The second term in (\ref{eq:defS}) is a function of $t$, 
      so (\ref{eq:defF}) becomes 
      \begin{equation}
        F = \frac{1}{2m}\frac{\partial^{2}S}{\partial q^{2}}
          = \frac{1}{2m}\frac{\partial^{2}W}{\partial q^{2}}. 
      \label{eq:defF2}
      \end{equation}
\item Next, putting $F$ back into (\ref{eq:defS}) and integrating 
      the second term, we derive the function $S$. 
\item Accordingly, putting this $S$ back into (\ref{eq:solu}), 
      we obtain the solution to the Schr\"{o}dinger 
      equation (\ref{eq:se}). 
\end{enumerate}

Now, we apply this method to the generating function 
$W_{1}(q,Q,t)$ which appears in (\ref{eq:gf1}). 
It is easy to check that 
\begin{equation}
a(t) = -m \dot{b}(t)
\label{eq:const}
\end{equation}
is required in order for $W_{1}(q,Q,t)$ to satisfy the Hamilton-Jacobi 
equation (\ref{eq:HJ}), where $\dot{b}(t)$ describes 
the time derivative of $b(t)$. 
We have two other conditions but these have no 
bearing on the following discussion. 
Calculating the function $F$ from (\ref{eq:defF2}) 
with (\ref{eq:const})
\begin{equation}
F(t) = \frac{1}{2m}\frac{-a(t)}{b(t)} 
     = \frac{1}{2m}\frac{m\dot{b}(t)}{b(t)} 
     = \frac{d}{dt} \ln \sqrt{b(t)}, 
\end{equation}
we have the exponential factor $S(q,t)$ 
\begin{equation}
S(q,t) = W_{1}(q,Q,t) + i \hbar \ln \sqrt{b(t)} .
\end{equation}

Thus, the solution of the Schr\"{o}dinger equation is 
\begin{equation}
\psi = \exp\left[ \frac{i}{\hbar} S(q,t)\right]
     = \sqrt{\frac{1}{b(t)}} 
       \exp\left[ \frac{i}{\hbar} W_{1}(q,Q,t) \right] .
\end{equation}
This is identical to eq.(\ref{eq:trans}) without the trivial 
constant in the square root 
which is obtained from the normalization of the wave function. 

This is the second result in this paper. We assume the solution 
of the Schr\"{o}dinger equation to be like (\ref{eq:solu}). Part of 
the function $S$ is the generating function $W_{1}$ which 
satisfies the Hamilton-Jacobi equation. Once we have 
the generating function which satisfies the Hamilton-Jacobi 
equation, then the solution of the Schr\"{o}dinger equation is 
calculated according to the procedure stated above. 

\section{ Applications to simple systems }

In this section, we will apply the method to two ideal systems. 

\subsection{Free particle}

The Hamiltonian is 
\begin{equation}
  H = \frac{p^{2}}{2m}, 
\end{equation}
where $m$ and $p$ are the mass and momentum of the particle. 

We consider the canonical transformation 
\begin{equation}
  \left(
    \begin{array}{@{\,}cc@{\,}}
        a(t) & b(t) \\
        c(t) & d(t)  
    \end{array}
  \right)
  =
  \left(
    \begin{array}{@{\,}cc@{\,}}
        1 & -\frac{t}{m} \\
        0 & 1  
    \end{array}
  \right), 
\label{eq:ct00}
\end{equation}
whose physical meaning is the Galilean transformation. 
The constraints (\ref{eq:pb}) and (\ref{eq:const}) are satisfied 
by these coefficients. 
The generating function (\ref{eq:gf1}) that results in the above 
transformation is 
\begin{equation}
W_{1}(q,Q,t) = -\frac{m}{t}qQ+\frac{m}{2t}q^{2}+\frac{m}{2t}Q^{2}
             = \frac{m}{2t}(q-Q)^{2}. 
\end{equation}
Since this is the Galilean transformation, the transformed 
Hamiltonian $K = H + \frac{\partial W_{1}}{\partial t}$ must 
be zero. So, $W_{1}$ also satisfies the Hamilton-Jacobi equation.

From the above canonical transformation, we obtain the 
transformation function or the solution to the Schr\"{o}dinger equation 
\begin{eqnarray}
  \bra q|Q;t\ket &=& \exp\left[\frac{i}{\hbar} S \right] \\ \nonumber
                 &=& \sqrt{\frac{m}{2\pi i \hbar t}}
                     \exp\left[\frac{i}{\hbar} 
                               \frac{m}{2t}(q-Q)^{2} \right] .
\end{eqnarray}
It is interesting that this function is identical to the 
Feynman propagator in the free Hamiltonian.

\subsection{Harmonic Oscillator}

The Hamiltonian is 
\begin{equation}
  H = \frac{p^{2}}{2m} + \frac{m\omega^{2}}{2}q^{2}, 
\end{equation}
where $m$, $p$ and $\omega$ are the mass, momentum and frequency of 
the particle.

We consider the canonical transformation 
\begin{equation}
  \left(
    \begin{array}{@{\,}cc@{\,}}
        a(t) & b(t) \\
        c(t) & d(t)  
    \end{array}
  \right)
  =
  \left(
    \begin{array}{@{\,}cc@{\,}}
        \cos \omega t & -\frac{1}{m\omega}\sin \omega t \\
        m\omega \sin \omega t & \cos \omega t  
    \end{array}
  \right), 
\label{eq:ctHO}
\end{equation}
whose physical meaning is a rotation in phase space. 
The constraints (\ref{eq:pb}) and (\ref{eq:const}) are satisfied 
by these coefficients. 
The generating function (\ref{eq:gf1}) that causes the above 
transformation is 
\begin{equation}
W_{1}(q,Q,t) = - \frac{m \omega}{\sin \omega t}qQ 
               + \frac{m\omega}{2 \tan \omega t}
                        \left(q^{2}+Q^{2}\right). 
\end{equation}
Since this is a rotation in phase space, the transformed 
Hamiltonian $K = H + \frac{\partial W_{1}}{\partial t}$ must 
be zero. So, $W_{1}$ also satisfies the Hamilton-Jacobi equation.

From the above canonical transformation, we obtain the 
transformation function or the solution to the Schr\"{o}dinger equation 
\begin{eqnarray}
  \bra q|Q;t\ket &=& \exp\left[\frac{i}{\hbar} S \right] \\ \nonumber
                 &=& \sqrt{\frac{m\omega}
                                {2\pi i \hbar \sin \omega t}}
                     \exp\left[\frac{i}{\hbar} 
                          \left\{ - \frac{m \omega}
                                     {\sin \omega t}qQ 
               + \frac{m\omega}{2 \tan \omega t}
                        \left(q^{2}+Q^{2}\right) \right\}
                                \right] .
\end{eqnarray}
It is interesting that this function is identical to the 
Feynman propagator of the harmonic oscillator.

\section{ Discussion }

We have investigated two methods which construct solutions to 
the Schr\"{o}dinger equation from classical canonical 
transformations. 
For linear canonical transformations in classical mechanics, 
we have constructed the quantum counterpart and this is related to 
the Hamilton-Jacobi theory. 

However, our method is restricted only to linear canonical 
transformations. 
There is a variety of canonical 
transformations in classical mechanics. 
For example, the generating function 
\begin{equation}
W_{1}(q,Q,t) = \frac{m\omega q^{2}}{2 \tan Q}
\end{equation}
is a good choice for obtaining a solution to the classical 
harmonic oscillator, 
while a naive application of Dirac's (\ref{eq:type1}) does not 
work in quantum mechanics~\cite{lee}.

\vspace{1cm}


\end{document}